\documentclass[reprint,preprintnumbers,amsmath,amssymb,aps,prd,tightenlines,longbibliography,nofootinbib,balancelastpage]{revtex4-2}

\usepackage{graphicx}
\usepackage{comment}
\usepackage[dvipsnames]{xcolor}
\usepackage[sort&compress]{natbib}
\usepackage{amsmath,amssymb,bm,bbm,slashed,subdepth}
\usepackage{soul}
\usepackage{xr-hyper}
\usepackage[colorlinks=true
,urlcolor=blue
,anchorcolor=blue
,citecolor=blue
,filecolor=blue
,linkcolor=red
,menucolor=blue
,linktocpage=true
,pdfproducer=medialab
,pdfa=true
]{hyperref}
\usepackage{cleveref}
\usepackage{enumerate}
\usepackage{epsfig, subfigure}
\usepackage{setspace}
\usepackage{booktabs, tabularx}
\usepackage{units}
\usepackage{placeins}
\usepackage{multirow}
\usepackage{mathtools}
\usepackage[normalem]{ulem}

\newcommand{\be}{\begin{eqnarray}}
\newcommand{\ee}{\end{eqnarray}}

\usepackage[page,toc,titletoc,title]{appendix}

\begin{document}

\preprint{CERN-TH-2026-184}

\title{Halbach Magnetic Weber Bars }

\author{Valerie~Domcke}
\author{Itay M. Bloch}
\affiliation{Theoretical Physics Department, CERN, 1 Esplanade des Particules, CH-1211 Geneva 23, Switzerland}

\begin{abstract}
Magnetic Weber Bars have been proposed to search for gravitational waves\textbf{} in the kHz to GHz regime by exploiting the mechanical deformation of a large magnet induced by a gravitational wave. Here we propose to increase the effectiveness of such devices by considering magnetic field configurations with strong gradients, albeit lower field strengths, such as Halbach arrays. We focus on {one of} the most challenging but most realistic signals, with short duration and low coherence, exploiting the ring-down period of the mechanical resonator. We show that with demonstrated technology this setup can reach sensitivities of $S_h^{1/2} \simeq 10^{-21}/\sqrt{\text{Hz}}$  at a broad set of frequencies around several resonance peaks at $\sim 10$~kHz, and $S_h^{1/2} \simeq 5 \cdot 10^{-20}/\sqrt{\text{Hz}}$ in a broadband search at higher frequencies. We discuss plausible upgrades to reach  $S_h^{1/2} \simeq (10^{-23} - 10^{-21})/\sqrt{\text{Hz}}$ in a broadband search covering 10~kHz - MHz.
\end{abstract}

\maketitle

\section{Introduction}

Historically, one of the earliest ideas for the measurement of gravitational waves (GWs) was the use of high-quality mechanical resonators, with the earliest efforts being so-called Weber Bars~\cite{Weber:1967jye}. In such sensors, a passing GW excites mechanical oscillations of a macroscopic object, which are then read out by a displacement sensor.
Over the past decades, more advanced resonant mass antenna concepts have searched for GWs in the kHz~\cite{hamilton1990resonant,Vinante:2006uk,Gottardi:2007zn,DaSilvaCosta:2014trv} and MHz~\cite{Goryachev:2021zzn,Campbell:2023qbf,Campbell:2025mks} regime, with further proposals extending both the frequency and sensitivity range~\cite{PhysRevD.54.2409,Ballantini:2005am,Gottardi:2006gn,Gottardi:2007zn,Goryachev:2014yra,Campbell:2023qbf,Aggarwal:2020umq,Berlin:2023grv,Domcke:2024mfu,Carney:2024zzk}, see Ref.~\cite{Aggarwal:2025noe} for a review.
To date, no credible GW signal has been measured by a resonant mass antenna,
yet the basic premise is of great interest, since it can be implemented in relatively small-scale experiments compared to current interferometry techniques~\cite{LIGOScientific:2014pky,VIRGO:2014yos,KAGRA:2020tym}.

Among the more recent proposed implementations, the ``Magnetic Weber Bar'' (MWB) proposal of Ref.~\cite{Domcke:2024mfu} pointed out that by using magnets as the resonant mass, one can enhance the potential sensitivity of such devices.
In that proposal, the GW mechanically deforms the current-carrying structure of a DC magnet, generating a small AC magnetic field component that can be read out with a pickup loop and a SQUID. The MWB combines the favorable mechanical response of a resonant mass with an efficient electromagnetic readout channel.

In this work, we explore an upgraded version of the MWB concept. Our configuration places magnetic sensors on a large resonant mass, and surrounds the mass with magnetic fields of $\sim$ Tesla scale amplitude, but extremely high gradients. Such configurations can be achieved over long distances using Halbach configurations~\cite{Mallinson:1973,Halbach:1980,Blumler:2023qvb}, and similar other alternatives. They leverage a special planar arrangement of permanent magnets to achieve high field gradients close to the surface. A deformation $\delta x$ of the resonant mass by a passing GW leads to a change in the measured magnetic field
\begin{equation}
\delta B \sim \frac{\partial B}{\partial x} \delta x \,.
\label{eq:dB-intro}
\end{equation}
For comparison, the proposal in Ref.~\cite{Domcke:2024mfu} assumed a solenoidal magnetic field with $\delta B/\delta x \sim B/L \sim 10$~T/m, with $L$ the characteristic scale of the magnet. Here, we can instead limit ourselves to modest magnetic fields and nevertheless boost this quantity by two orders of magnitude due to the gradient enhancement.
We refer to this idea as ``Halbach Magnetic Weber Bar'', in reference to earlier terminology, though our devices are not necessarily bar shaped.

The two fundamental noise components we have to contend with are thermomechanical noise and SQUID noise. Boosting the transfer function~\eqref{eq:dB-intro}  increases the signal to SQUID noise ratio. Conversely, the thermomechanical noise is enhanced by the same mechanism, leaving the signal to thermomechanical noise ratio unchanged. Consequently, our peak sensitivity near mechanical resonance frequencies, typically limited by thermomechanical noise, remains unaffected by the improved transfer function. However, the effective lowering of the SQUID noise floor implies that the bandwidth over which this peak sensitivity can be achieved is significantly broader. Moreover, at higher frequencies, where the sensitivity is typically SQUID-noise limited and already broadband, the improved transfer function directly enhances the strain sensitivity across the band.

In addition to presenting this idea for an upgraded design, we furthermore extend the single-mechanical-mode simplified picture used in Ref.~\cite{Domcke:2024mfu}, computing the sensitivity in the more realistic case of multiple mechanical resonances.

To illustrate the expected shape of GW signals, we carry out a brief analysis in the time domain of the system response, in addition to the more common frequency-domain analysis. This demonstrates an aspect that is usually not emphasized, which is that the long lifetime of mechanical modes implies that the signal of a transient GW event would in fact mostly be measured during the relaxation of the mode, long after the GW has passed.
This simplifies the discussion of the response of the system to a GW, as one can neglect the motion of the parts of the setup which don't respond resonantly to the GW.

The remainder of this paper is organized as follows. In Sec.~\ref{sec:GWantennas} we review the response of a resonant mass antenna to a passing GW. Section~\ref{sec:setup} details the concrete setup we have in mind, featuring a resonant sphere instrumented with multiple pickup loops, surrounded by a magnetic configuration with strong field gradients. The dominant noise sources and the resulting projections for the sensitivity are given in Sec.~\ref{sec:sensitivity}, before concluding.  In the appendices we give additional information on a variety of more technical calculations, including a more detailed treatment of the mechanical resonances of a solid sphere, further description of the magnetic field of the array, a time domain analysis of a transient toy-signal, and the extrapolation of our sensitivity to higher frequencies.
The code and data needed to reproduce all plots, tables, and other computations are provided in Ref.~\cite{code-ref}.

\section{Resonant Mass GW Antennas}
\label{sec:GWantennas}
The mechanical response of a resonant sphere to a passing GW is a well-studied problem in the context of GW searches in the kHz regime~\cite{Forward:1971mel,DaSilvaCosta:2014trv,Gottardi:2007zn}. Here we briefly review the key aspects needed for our discussion, referring to App.~\ref{app:modes} and Ref.~\cite{Lobo:1995sc} for a more detailed derivation.

We start from the equation of motion for the displacement field ${\bm u}(\bm x, t)$ of an elastic material,
\begin{equation}
 \rho \partial_t^2 {\bm u} = (\lambda + \mu) {\bm \nabla} ( {\bm \nabla} \cdot {\bm u}) + \mu {\bm \nabla}^2 \bm u - \rho \hat \Gamma \partial_t {\bm u}  + \bm f^\text{ext}\,,
 \label{eq:eom}
\end{equation}
where $\rho$ denotes the density, $\lambda$ and $\mu$ are the Lam\'e coefficients, and $\hat \Gamma$ is a damping term.
The external force density $\bm f^\text{ext}$ is induced by the GW,
\begin{align}
 f^\text{ext}_i = \frac{\rho}{2} \ddot h_{ij} x^j \,.
 \label{eq:fext}
\end{align}
For the purpose of our discussion we can restrict ourselves to the low-frequency regime,  $\omega_g \ll 1/R$, with $\omega_g$ the GW frequency and $R$ the sphere radius. It is then convenient to work in the proper detector frame, in which the GW acts as the Newtonian force given by Eq.~\eqref{eq:fext}, where $h_{ij}$ is the GW in the transverse traceless frame~\cite{Maggiore:2007ulw}.

Equation~\eqref{eq:eom} can be solved by the ansatz
\begin{align}
 \bm u(\bm x,t) = \sum_{\bm n} c_{\bm n}(t) \, \bm u_{\bm n}(\bm x) \,,
 \label{eq:u_ansatz}
\end{align}
with $\bm u_{\bm n}$ a complete set of energy eigenmodes normalized as
\begin{align}
 \int_V d^3x \,\bm u_{\bm m}^*\cdot  \bm u_{\bm n} = V \delta_{\bm n, \bm m} \,,
 \label{eq:norm}
\end{align}
and the damping term is taken to be diagonal in this basis,
\begin{align}
 \hat \Gamma \bm u_{\bm n} = \frac{\omega_{\bm n}}{Q_{\bm n}} \bm u_{\bm n}\,.
\end{align}
Here $V$ is the volume of the sphere, and $\omega_{\bm n}$ and $Q_{\bm n}$ denote the eigenfrequency and quality factor of the mode $\bm n$. For a solid sphere these eigenmodes are known analytically. 
They can be classified in toroidal and spheroidal modes, and obey free boundary conditions $\sigma_{ij} (\hat e_r)_j = 0$ at $r = R$, where $\sigma_{ij}$ denotes the stress tensor. It is useful to separate the angular and radial dependence of the modes of interest using spherical harmonics, so that  $\bm n = \{n, l, m\}$ with $\{l,m\}$ the spherical harmonic indices, and the coefficients are given by spherical Bessel functions which depend on the eigenfrequency $\omega_{nl} \sim n v_s/R$ (up to an $l$-dependent offset) with $v_s$ the speed of sound, see  Ref.~\cite{Lobo:1995sc} and  App.~\ref{app:modes} for details.

The response to the GW is encoded in the coefficients $c_{\bm n}(t)$. Performing a mode and Fourier expansion of Eq.~\eqref{eq:eom} and multiplying with $\bm u_{\bm n}^*$, one obtains
\begin{align}
 \tilde c_{\bm n}(\omega) & = \frac{\int_V d^3r \, \bm u_{\bm n}^* \cdot \tilde{ \bm f}^\text{ext}({\bm r}, \omega) }{ \rho V (\omega_{\bm n}^2 - \omega^2 - i \omega \omega_{\bm n}/Q_{\bm n})}  \nonumber \\
 & =  R \, \eta^A_{\bm n} \, \widetilde{\ddot{h}}_A \, (\omega_{\bm n}^2 - \omega^2 - i \omega \omega_{\bm n}/Q_{\bm n})^{-1} \,,
 \label{eq:cn-general}
\end{align}
where tilde denotes Fourier transformed quantities and we have exploited that in the low-frequency regime,  $h_{ij}({\bm x},t) \simeq h_A(t) e_{ij}^A$. Here $e_{ij}^A(\hat{\bm k})$ is the polarization tensor with respect to the GW propagation direction $\hat{\bm k}$, defined in App.~\ref{app:conventions}, and ${\eta}_{\bm n}$ is the overlap factor,
\begin{align}
 \eta^A_{\bm n} = \frac{1}{2 R V} \int_V d^3x \, e_{ij}^A(\hat{\bm k}) x^j (u_{\bm n}^*)^i \,.
\end{align}
For the case of a sphere, non-vanishing overlap factors are found only for spheroidal modes with $l = 2$, reflecting the tensor nature of the GW. The overlap function is maximized when the orientation $m$ is optimally aligned with the GW, and is then approximately $\eta^A_{\bm n} \simeq 0.2/n^2$.

The very small number of non-vanishing overlap factors has several implications. It will help in distinguishing signal from noise, since thermomechanical noise excites all modes whereas the signal is concentrated in the $l=2$ modes.
A suitable readout geometry will moreover allow to distinguish the vibrations in different $l=2$ modes and hence be able to reconstruct the sky position of the GW source and polarization of the GW.

In the following we will focus for simplicity on a GW with $\times$ polarization coming from the z-direction, which couples exclusively to the $l = 2$, $m =\pm2$ spheroidal modes of the sphere, out of which we focus for presentation purposes on the $m=2$ mode. The extrema of the displacement on the surface $|\bm u_{n22}(R)|$ occur along the equator and consist of a radial and angular motion (in the direction of the azimuthal angle $\phi$). The latter peaks at multiples of $\phi = \pi/2$, while the peaks of the former are shifted by $\pi/4$, implying that a node in radial directions coincides with a maximal displacement in the angular direction and vice versa. The maxima of the {dimensionless radial surface amplitudes $\alpha_{n22}(\theta,\phi)=\hat e_r\cdot\bm u_{n22}(R,\theta,\phi)$ of the normalized eigenmodes} are typically of magnitude ${\cal O}(0.1 - 1)$. The setup described below will primarily target the radial motion, though similar ideas can also be exploited to read out the angular motion.\footnote{The eigenmodes of the sphere feature large angular gradients for high $m$ modes, which would substantially contribute to thermomechanical noise. This motivates our choice of readout system which is optimized for radial modes.}

\section{Setup and GW response}
\label{sec:setup}
Halbach arrays, most commonly known from their use in fridge magnets, are constructed from a special arrangement of permanent magnets which creates a high gradient magnetic field on one side of the array while nearly cancelling it on the other (see ~\cite{Mallinson:1973,Halbach:1980,Blumler:2023qvb} and App.~\ref{app:halbach}). More broadly, a variety of magnetic geometries exist which can lead to high magnetic gradients, and can be used in a variety of applications, including within particle physics~\cite{asparuhov:tel-04207215}. Throughout this work, we will refer to our configuration as a Halbach array, despite not truly needing the cancellation of magnetic fields on the far side of the magnets, and thus possibly being amenable to other geometries as well.
\begin{figure}
 \centering
 \includegraphics[width = \columnwidth]{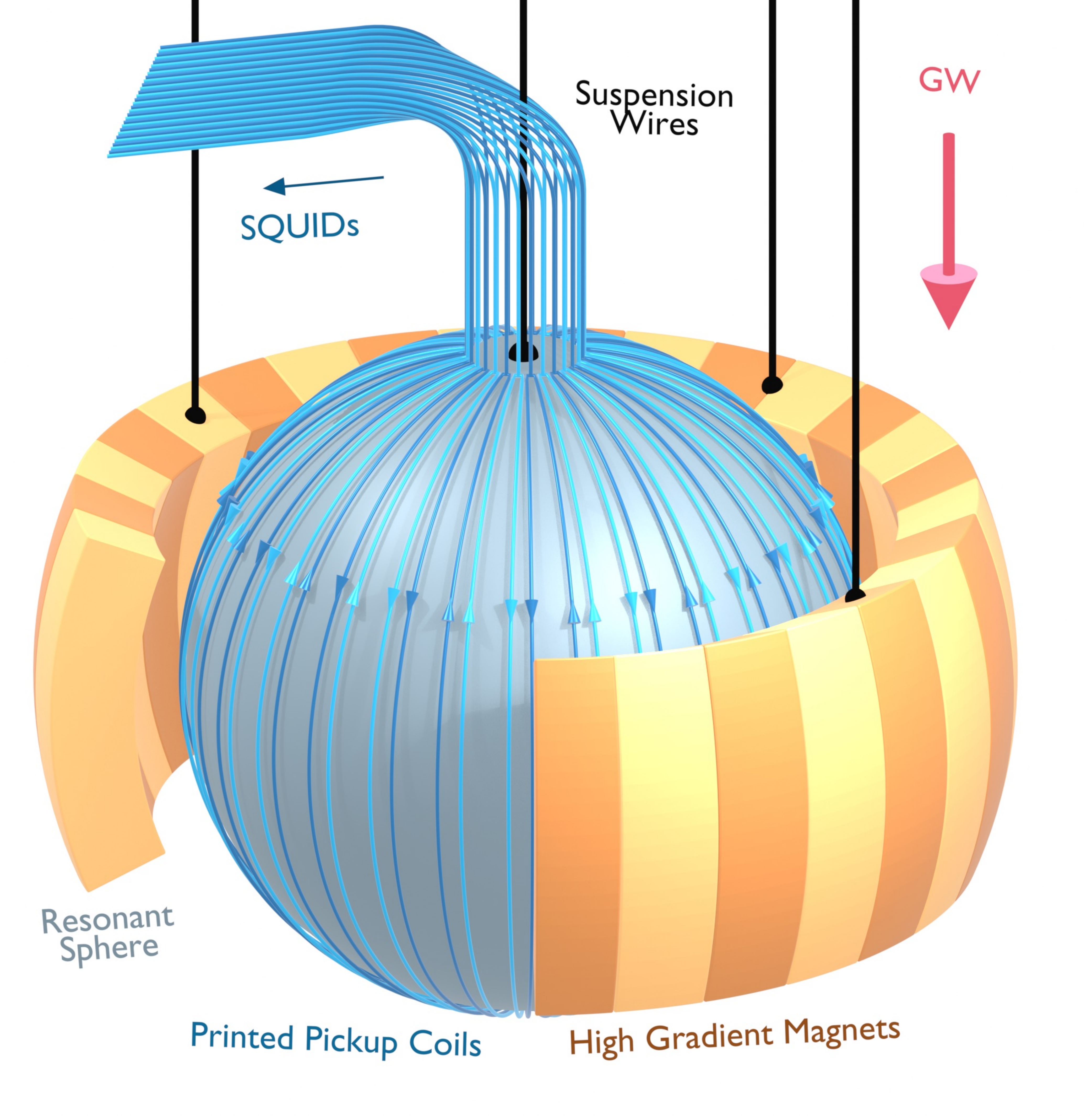}
 \caption{Schematic of the proposed setup. Meridional pickup loops are printed or otherwise rigidly connected to the resonant sphere. Neighboring loops alternate between cyan and blue to help visually distinguish the separate loops. Arrowheads indicate the winding direction on each trace, and each loop is centred on one period of the surrounding high-gradient Halbach array (orange and yellow). Multiple loops are combined into a few pickup circuits connected to individual SQUID readouts. The red arrow indicates an incoming $\times$-polarized GW coming from the $z$ direction. The GW causes the sphere to vibrate as it passes, and
 the sphere continues to ring down for a much longer time scale, controlled by the quality factors of its modes. Conversely, the motion of the magnet support is assumed to be quickly damped. The strong field gradient converts small displacements of the sphere surface into large changes in the magnetic flux through the loops.}
 \label{fig:setup}
\end{figure}

In our setup, we place such magnets just outside a resonant sphere, and imprint the resonant sphere's surface with pickup loops to detect any change in flux due to the relative motion of sphere and magnets. The passing of the GW will lead to a Newtonian force on both the sphere and the magnet, and in principle one needs to account for both~\cite{Domcke:2024mfu}. Here, however, we make use of the fact that most high-frequency GW signals we are interested in are of short duration ${\cal O}(1/\omega_g)$ and that the resonant sphere is an excellent resonator, cf.~\ref{app:time-domain}, while we imagine the magnets to be mounted on a structure that quickly dampens any oscillation in this frequency range. Hence up to negligible corrections of order $1/Q$, it suffices to consider the response of the resonant sphere.\footnote{If needed, these could be included by computing the full transfer function as in \cite{Domcke:2024mfu} or by calibration of the response function with controlled mechanical perturbations.}
Hence, within this approximation, the magnets are treated as static in the proper detector frame and we can estimate the change in the magnetic flux through the pickup loop as
\begin{align}
  \Phi_h \simeq \sum_n c_{n22}(t)\,
 \langle A\alpha_{n22}\partial_r B_r\rangle\,,
  \label{eq:Phi}
\end{align}
with the effective instrumented area $A$ taken as $A=4\pi R^2/\sqrt{2}$ after coherently combining the four quadrants (see Appendices~\ref{app:modes} and~\ref{app:halbach}), $\partial_r B_r$ the magnitude of the gradient of the radial component of the magnetic field of the Halbach array at its extrema, and we recall $c_{n22}(t) \sim \eta_{n22} h_0^\times R$.

To obtain the optimal signal strength, one can aim for large overall magnetic field strengths or for locally large gradients. In the original MWB proposal~\cite{Domcke:2024mfu}, a superconducting solenoidal magnet was considered, with $B' \sim B_0/R$ where $R$ is a characteristic scale of the setup. Here, we aim for peak field strengths of $B_0 \simeq 0.5-1 ~{\rm T}$, much below the $16~T$ considered there, but compensate for this by locally strong gradients, $B' \simeq {\rm T}/{\rm mm}$  over a distance of ${\rm mm}$. Similar, though $\mathcal{O}(3)$ lower gradients have been demonstrated in macroscopic undulators~\cite{asparuhov:tel-04207215}. It seems likely that by reducing the required precision in the magnetic field profile, the targeted performance would become viable, and even higher magnetic gradients might be possible.

Our basic setup is shown in Fig.~\ref{fig:setup}. The alternating winding directions match the alternating magnetic-field gradient so that the displacement-induced fluxes add coherently. In practice, we envisage approximately $5$--$20$ independent pickup circuits, each connected to a SQUID, to
retain the spatial information needed to reconstruct several $n2m$ modes. Our approach below simplifies the description by taking a single circuit, with further details given in App.~\ref{app:halbach}.

We can now estimate the power spectral density (PSD) of the flux induced by the GW as seen by the pickup loop.
From the response for a single mode we obtain
\begin{align}
 S^\text{sig}_{\Phi,\bm n}(\omega) =
|\langle A\alpha_{\bm n}\partial_r B_r\rangle|^2 S^\text{sig}_{x,\bm n}(\omega)\,.
 \label{eq:SPhi_sig}
\end{align}
Here $S^\text{sig}_{x,\bm n}(\omega)=\tilde c_{\bm n}(\omega)\tilde c_{\bm n}^*(\omega)/\tau$ denotes the mechanical-vibration PSD of mode $\bm n$. It quantifies the readout-independent excitation of the mechanical mode, while the transfer function based on weighted averaging $\langle A\alpha_{\bm n}\partial_r B_r\rangle$ converts this vibration into magnetic flux.  Here $\tau$ is a time scale set by the instrument/analysis over which we search for excess power, which will drop out in our results for the strain sensitivity.
Inserting Eq.~\eqref{eq:cn-general} yields
\begin{align}
 S^\text{sig}_{x,\bm n}(\omega)=\frac{(\eta_{\bm n}^\times R)^2\omega^4}{(\omega^2-\omega_{\bm n}^2)^2+(\omega_{\bm n}\omega/Q_{\bm n})^2}S_h(\omega) \,.
\end{align}
with $ S_h(\omega) = \tilde h(\omega) \tilde h^*(\omega)/\tau$.

To consider contributions from different modes, we note that at any given frequency, the total PSD will be largely dominated by a single mode, and hence to good approximation $S_\Phi^\text{sig} \simeq \sum_n S_{\Phi, n}^\text{sig}$. Due to the $1/n^2$ scaling of the overlap factor, the sensitivity is largest near low $n$ modes.

\section{Noise Sources and Sensitivity}
\label{sec:sensitivity}

We focus on two fundamental noise sources, thermomechanical noise and SQUID noise. As in~\cite{Domcke:2024mfu} we take the corresponding PSDs for the noise induced flux as seen by the pickup loop to be
\begin{align}
 S_{\Phi,\bm n}^\text{th.mech.}(\omega)
 &=
 \frac{ |\langle A\alpha_{\bm n}\partial_rB_r\rangle|^2 \, 2T\omega_{\bm n}/(Q_{\bm n}M)}{(\omega^2-\omega_{\bm n}^2)^2+(\omega_{\bm n}\omega/Q_{\bm n})^2} \,\,, \\
 S_\Phi^\text{SQUID}(\omega) & = \kappa^{-2} 10^{-12} \Phi_0^2/\text{Hz}  \,,
\end{align}
where  $M$ and $T$  are the mass and temperature of the resonant sphere, $\kappa \simeq 0.002$ arises from the inductive coupling between the pickup loop and the SQUID, see App.~\ref{app:halbach}, and $\Phi_0 = \pi \hbar/e = 2.07 \times 10^{-15}$~Wb is the flux quantum.
We assume the surrounding magnets are mounted on some very massive structure, so that we can neglect their thermomechanical noise.

As for the signal PSD, we account for thermomechanical noise of different modes by summing $S_{\Phi}^\text{th.mech.} = \sum_{\bm n} S_{\Phi, \bm n}^\text{th.mech.}$. We will assume that using frequency and spatial information we can select only the contributions from the $\bm n = n22$ modes. In App.~\ref{app:noise} we instead include the contributions of all modes to the thermomechanical noise around the first three $n22$ resonances. This shows that on the $n22$ resonances, their contribution is small, while off-resonance, the sensitivity is typically limited by SQUID noise and hence again, the thermomechanical noise contribution from modes other than the $n22$ mode is largely irrelevant. See also Fig.~\ref{fig:sensitivity-noise-all-modes} in App.~\ref{app:noise}.
The filtering and identification of different modes has been demonstrated in the vicinity of the $122$ mode in classical resonant mass antennas such as MiniGRAIL~\cite{Gottardi:2006gn,Gottardi:2007zn}.

A GW signal is detectable if the signal to noise ratio,
\begin{align}
 \text{SNR}^2 \simeq  \frac{t_\text{int}}{\pi} \int_0^\infty d\omega \left( \frac{S_\Phi^\text{sig}(\omega)}{S_\Phi^\text{th.mech.}+S_\Phi^\text{SQUID}} \right) \,,
 \label{eq:SNR}
\end{align}
exceeds some ${\cal O}(10)$ threshold value, where $t_\text{int}$ is the integration time. To display this as a sensitivity to gravitational wave strain, we introduce the strain-equivalent noise,
\begin{align}
 S_h^\text{noise}(\omega) = \frac{S_\Phi^\text{th.mech.}+S_\Phi^\text{SQUID}}{(S_\Phi^\text{sig}/S_h)} \,,
\end{align}
which indicates the GW strain $S_h(\omega)$ for which the bracket in Eq.~\eqref{eq:SNR} is unity.

\begin{figure*}
 \centering
 \includegraphics[width = 0.7 \textwidth]{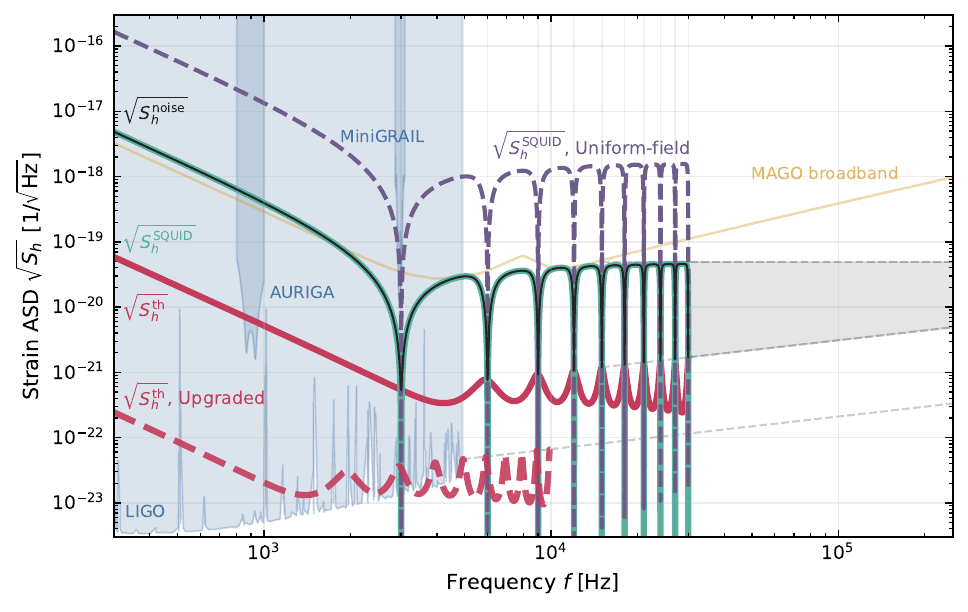}
 \caption{Strain-equivalent noise amplitude spectral density for a resonant sphere with $R = 0.34$~m, $B'_r = 1~{\rm T/mm}$, and $Q_n = Q =  10^7$, including the lowest 10 $n22$ modes. The total noise is shown in black, with thermomechanical and SQUID contributions in solid red and green, respectively. For reference, the dashed purple curve gives the SQUID strain-equivalent noise for the uniform-field estimate $B'_r = B_0/R$ with $B_0 = 10$~T, and the dashed red curve gives the projected thermomechanical-noise floor for the upgraded configuration described in the text. Published LIGO~\cite{LIGO:2024kkz}, AURIGA~\cite{Vinante:2006uk}, and MiniGRAIL~\cite{Gottardi:2007zn} sensitivities are shown in pale blue, while the gold curve shows the broadband projection for the proposed MAGO 2.0 experiment~\cite{Berlin:2023grv}.
 In dashed gray we show high-frequency extrapolations of the respective sensitivity curves, with the total noise expected to lie within the shaded gray region, see App.~\ref{app:noise}. }
 \label{fig:sensitivity}
\end{figure*}

We show the resulting sensitivity in terms of the amplitude spectral density (ASD), which is the square root of the PSD, as function of linear frequency $f = \omega/(2 \pi)$ in Fig.~\ref{fig:sensitivity}. Our main result is depicted as the black curve,  assuming a magnetic field gradient of $B'_r = 1~{\rm T/mm}$, and simplified benchmark parameters inspired by MiniGRAIL~\cite{Gottardi:2007zn}: $R = 0.34$~m, $T = 4.2$~K, $M = 1.3 \times 10^3$~kg, ${Q_{n22}} = Q = 10^7$,
$f_1 = \omega_1/(2 \pi) = 3$~kHz,
${\omega_{n22}} = n \omega_1$, ${\langle A\alpha_{n22}\partial_rB_r\rangle=0.3\,A B'_r}$, and ${\eta_{n22}^\times} = 0.2/n^2$, including the lowest ten $n22$ modes. A calculation using the numerical mode spectrum and mode-dependent coefficients of a sphere is given in App.~\ref{app:modes} and yields a similar sensitivity.

Across most of the plotted frequency range, the total sensitivity (black) is limited by the SQUID contribution (green). It can therefore be improved by reducing the SQUID noise or by enhancing the transfer function in Eq.~\eqref{eq:SPhi_sig}, for example by increasing the sphere radius and hence effective readout area or by using magnetic-field configurations with even larger gradients. The benchmark values used in Fig.~\ref{fig:sensitivity} are based on readily available technology.

In the immediate vicinity of each $n22$ mechanical resonance, the sensitivity is limited by thermomechanical noise (solid red curve). The displacement-to-flux transfer function is the same for thermomechanical noise and the signal, so a stronger magnetic-field gradient does not improve the peak sensitivity in this regime. It does, however, broaden the frequency interval over which thermomechanical noise dominates, allowing the detector to approach its peak sensitivity over a wider band.

For reference, the dashed purple curve labeled ``Uniform-field'' in Fig.~\ref{fig:sensitivity} shows the SQUID-limited strain noise obtained without gradient enhancement, i.e. using $B'_r = B_0/R$ with $B_0 = 10$~T. This corresponds to the parametric scaling considered in Ref.~\cite{Domcke:2024mfu}.\footnote{In particular, the more optimistic curve in Ref.~\cite{Domcke:2024mfu} assumed a mass of 40~t, while the geometry of a hollow cylinder allowed for a length of 4~m and a pickup loop around the end cap with radius 3~m. Moreover, we find the values for $\alpha$ and $\eta$ to be slightly smaller in the spherical case considered here. See also App.~\ref{app:modes}.}
In addition, the dashed red curve shows the thermomechanical-noise floor of a possible upgraded instrument, assuming that SQUID noise can be suppressed sufficiently. Its parameters follow an improved MiniGRAIL configuration described as achievable with the ``current technology'' of 2007~\cite{Gottardi:2006gn,Gottardi:2007zn}, in particular $T = 50$~mK and $R = 1$~m, which gives  $f_1 =  987$~Hz. The remaining parameters are as above.

A further conceivable improvement is the implementation of a resonant LC circuit coupled to the  pickup loop, similar to what was envisaged e.g.\ for DMRadio-GUT~\cite{DMRadio:2022jfv} and in Ref.~\cite{Domcke:2024mfu}. While this would not reduce the thermomechanical noise (which again would be enhanced in the same way as the signal), this approach could be taken to reduce the SQUID noise, at the cost of compromising the broadband sensitivity. The broader frequency range could then be recovered by either a scanning strategy, or the use of many readout circuits sensitive to different frequencies.

While the main focus of this paper is on transient signals, the setup is also sensitive to a stochastic gravitational wave background (SGWB). However, even the optimistic thermomechanical-noise floor shown by the dashed red curve in Fig.~\ref{fig:sensitivity} falls short by about a factor 30 of reaching the BBN limit on the energy density in GWs, $\Omega_\text{gw} h^2 < 10^{-6}$, with $\rho_\text{gw} = \rho_c \int d\ln f\, \Omega_\text{gw}(f)$.

What would it take to reach to the cosmologically highly motivated region below this limit? To illustrate the challenges, we give one example: A hollow instead of solid resonator leads to larger displacements at constant total mass due to its larger size. This is evident in the $1/(\eta R)^2$ scaling of $S_h^\text{noise}$ in the thermal noise limited regime. Moreover, a larger size leads to lower resonance frequencies, which gives a better energy reach at given strain reach, since $\rho_\text{gw} \sim h^2 \omega^2 M_P^2$. Following these arguments, we find that the BBN limit can be reached at about a kHz (marginally improving over the sensitivity of LIGO in this range) for the most optimistic parameter choice considered in Ref.~\cite{Coccia:1997gy}: a sphere of 2~m radius and 81~cm thickness with a weight of 200~tons, while keeping all other parameters as in the optimistic thermal limited case above. Clearly, the manufacturing (with very high quality factor), suspension and cooling of such an object is an extremely challenging engineering task.

\section{Conclusions}

Large magnetic field gradients can translate tiny mechanical motion induced by a passing GW into measurable electromagnetic signals read out by pickup loops coupled to SQUIDs. Magnetic Weber Bars combine this with the classical advantages of resonant mass antennas, i.e.\ a very large mass optimizing the coupling to GWs and a high mechanical quality factor, boosting the sensitivity on resonance. In this paper we further boost this efficiency by considering a setup with large magnetic field gradients. This allows us to reduce to moderate field strengths, thus facilitating the technical implementation.

In preparation of implementation of the Magnetic Weber Bar concept, we have also for the first time included a more realistic population of mechanical resonances and discussed several important data analysis aspects. Focusing on realistic GW signals, which typically are of short duration and have broad frequency spectra, we discuss spectral and spatial filtering to increase the signal-to-noise ratio. We argue that to good approximation, it suffices to concentrate on the ring down phase of the resonator, which significantly simplifies the response function.

The projected sensitivity based on previously demonstrated GW detector technology does not yet probe the parameter space of expected signals~\cite{Aggarwal:2025noe}. However, several avenues for future improvement are yet to be explored. The most obvious is to target more ambitious parameters in the existing setup, i.e.\ a lower temperature, a larger resonant mass, and stronger field gradients. Significant improvements seem possible by altering the shape of the resonator. The choice of a solid sphere in this work was merely for simplicity. On the contrary, a hollow object has the advantage of significantly enhancing the surface (and hence the readout region) while keeping the mass in a realm where suspension is mechanically feasible~\cite{Coccia:1997gy}. More moderate improvement factors can be gained by adapting the shape to reflect the symmetry of the GW, thus enhancing the overlap factor of the lowest modes to $\eta \sim 1$.

\vspace{0.2cm}
\noindent {\it Acknowledgements.}

We thank Sebastian Ellis, Joachim Kopp, Nick Rodd and Ben Safdi for useful discussions over the course of this project, and moreover Sebastian Ellis and Nick Rodd for feedback on the draft. VD acknowledges the support by the European Research Area (ERA) via the UNDARK project (project number 101159929).

\bibliographystyle{utphys}
\bibliography{ref}

\clearpage
\newpage
\maketitle
\onecolumngrid
\begin{center}
\textbf{\large Halbach Magnetic Weber Bars} \\
\vspace{0.05in}
{ \it \large Supplemental Material}\\ 

\end{center}
\setcounter{equation}{0}
\setcounter{figure}{0}
\setcounter{table}{0}
\setcounter{section}{0}
\renewcommand{\theequation}{S\arabic{equation}}
\renewcommand{\thefigure}{S\arabic{figure}}
\renewcommand{\thetable}{S\arabic{table}}
\renewcommand*{\thesection}{S.\Roman{section}}
\renewcommand*{\theHequation}{S.\arabic{equation}}
\renewcommand*{\theHfigure}{S.\arabic{figure}}
\renewcommand*{\theHtable}{S.\arabic{table}}
\renewcommand*{\theHsection}{S.\Roman{section}}
\interfootnotelinepenalty=10000

\setstretch{1.1}

In this Supplemental Material  we give additional information on a variety of more technical calculations, including a more detailed treatment of the mechanical resonances of a solid sphere, a more detailed description of our magnetic field profile and readout scheme, a time domain analysis of a transient toy-signal, and the extrapolation of our sensitivity curves to higher frequencies.

\section{Notation and conventions}
\label{app:conventions}
For a GW propagating in the direction
\begin{align}
 \hat{\bm k} = ( - \sin \theta_g \cos \phi_g, - \sin \theta_g \sin \phi_g, - \cos \theta_g ) \,,
\end{align}
(with the sign convention chosen so that $\theta_g, \phi_g$ label the sky position of the source in the usual way) we construct the transverse traceless tensors normalized to $ e_{ij, A}  e^{ij}_{B} = \delta_{A B}$ as
\begin{align}
 e^+_{ij} = \frac{1}{\sqrt{2}} (u_i u_j - v_i v_j ) \,, \quad e^\times_{ij} = \frac{1}{\sqrt{2}} ( u_i v_j + v_i u_j)\,,
\end{align}
with
\begin{align}
 \bm v = (\sin \phi_g, - \cos \phi_g, 0) \,, \quad \bm u = \bm v \times \hat{\bm k} \,.
\end{align}
The gravitational wave in transverse traceless gauge can then be expressed as
\begin{align}
 h_{ij}(\bm x,t) = (h_+ e^+_{ij} + h_\times e^\times_{ij} ) \exp[ i \omega_g(t - \hat{\bm k} \bm x) + i\varphi]  + h.c. \,.
\end{align}
In particular, for a GW propagating in the $z$-direction this yields
\begin{align}
 h_{ij}(\bm x) = \frac{1}{\sqrt{2}} \begin{pmatrix}
                                      h_+ & - h_\times & 0 \\
                                      - h_\times & h_+ & 0 \\
                                      0 & 0 & 0
                                    \end{pmatrix}  \exp[ i \omega_g  (t - z)] + h.c.\,.
\end{align}
For Fourier transformations, we use the symmetric convention
\begin{align*}
 x(t) = \int \frac{d \omega}{\sqrt{2 \pi}} \tilde x(\omega) e^{- i \omega t} \,, \quad \tilde x(\omega) = \int \frac{d t}{\sqrt{2 \pi}} x(t) e^{i \omega t}  \,.
\end{align*}
We use two-sided PSDs, satisfying $S(-f ) = S(f )$, so that the total power results after integration of the PSD over $f = [- \infty, \infty]$.

\section{Eigenmodes of a resonant sphere}
\label{app:modes}

In this appendix we review briefly the properties of the eigenmodes of a solid sphere, following Ref.~\cite{Lobo:1995sc}. We start with the equation of motion of the displacement field~\eqref{eq:eom}, setting the external force and the friction term to zero:
\begin{equation}
 \rho \partial_t^2 {\bm u} = (\lambda + \mu) {\bm \nabla} ( {\bm \nabla} \cdot {\bm u}) + \mu {\bm \nabla}^2 \bm u \,. 
 \label{eq:eom-free}
\end{equation}
We employ a Helmholtz decomposition into irrotational and gradient-free modes, which can be obtained from the Helmholtz potentials $\phi$ and $\psi$, respectively,
\begin{align}
  \bm u(\bm x) = \frac{C_0}{q^2} \bm \nabla \phi(\bm x) + \frac{i C_1}{k} \bm L \psi(x) + \frac{i C_2}{k^2} \bm \nabla \times \bm L \psi(x) \,,
\end{align}
where $C_{0,1,2}$ are constants that will be determined by the boundary conditions, $\bm L = - i \bm x \times \bm \nabla$ is the angular momentum operator, and $k$ and $q$ are the eigenvalues of the Helmholtz equations,
\begin{align}
 (\bm \nabla^2 + k^2) \psi(\bm x) = 0 \,, \quad (\bm \nabla^2 + q^2) \phi(\bm x) = 0 \,,
\end{align}
and are related to the angular eigenfrequency $\omega$ of the mode by the material properties,
\begin{align}
 k^2 = \frac{\rho \omega^2}{\mu} \,, \quad q^2 = \frac{\rho \omega^2}{\lambda + 2 \mu} \,.
\end{align}
The benchmark values used in the main text are motivated by the MiniGRAIL and Schenberg resonant sphere gravitational wave detectors~\cite{Gottardi:2007zn, Liccardo:2023nzv}, which used a solid sphere of CuAl6\%. Concretely, we will take $\rho = 8.0 \times 10^3~\text{kg/m}^3$, $\mu = 48.1$~GPa and $\lambda = 99.5$~GPa, which corresponds to sound speeds
of about $2.5-5.0$~km/s.
The solution to the Helmholtz equations is given by
\begin{align}
 \phi(\bm x) = j_l(q r) Y_{lm}(\hat e_r) \,, \quad \psi(\bm x) = j_l(k r) Y_{lm}(\hat e_r) \,,
\end{align}
with $Y_{lm}$ the spherical harmonic and $j_l(z)$ the spherical Bessel function.

To determine the coefficients $C_{0,1,2}$ and the eigenfrequency, we need to solve the boundary condition\footnote{
Ref.~\cite{Domcke:2024mfu} imposed weaker boundary conditions $\int_{\partial V} dS \, \sigma_{ij} \hat n_j = 0$ with $\hat{\bm n}$ the normal vector to the surface $\partial V$, in order to obtain analytical solutions for the mode functions. This should be taken into account when comparing the overlap factors $\eta$ in the two cases.
}
\begin{align}
 \sigma_{ij} \, (\hat e_r)_j = 0 \quad \text{at   } r = R \,,
 \label{eq:boundary-app}
\end{align}
with the stress tensor $\sigma_{ij} = \lambda u_{kk} \delta_{ij} + 2 \mu u_{ij}$ obtained from the strain tensor $u_{ij} = (\partial_i u_j + \partial_j u_i)/2$, which in turn is a function of the displacement field $\bm u$ which we wish to determine.\footnote{The use of the flat-space boundary conditions is justified in the low frequency regime, see~\cite{Gue:2026kga}.} The solutions fall in two classes, referred to as toroidal and spheroidal modes. In the following we focus on the spheroidal modes only, as the toroidal modes don't couple to our radial readout structure and moreover cannot be excited by GWs, as can be checked explicitly by the vanishing of the overlap factor~\cite{Lobo:1995sc}. The spheroidal modes feature $C_1 = 0$, but obtain contributions both from the irrotational and gradient free modes. They can be expressed as
\begin{align}
 \bm u_{nlm}(\bm x) = A_{nl}(r) Y_{lm}(\hat e_r) \hat e_r - i B_{nl}(r) \hat e_r \times \bm L Y_{lm}(\hat e_r) \,,
\end{align}
and solve the boundary conditions for
\begin{align}
 A_{nl}(r) & = {\cal N}_{nl} \left[ \beta_3(k_{nl} R) j'_l(q_{nl} r) - l (l + 1) \frac{q_{nl}}{k_{nl}} \beta_1(q_{nl} R) \frac{j_l(k_{nl} r)}{k_{nl} r} \right] \,, \\
B_{nl}(r) & = {\cal N}_{nl} \left[ \beta_3(k_{nl} R) \frac{j_l(q_{nl} r)}{q_{nl} r} -  \frac{q_{nl}}{k_{nl}} \beta_1(q_{nl} R) \frac{ \{ k_{nl} r j_l(k_{nl} r) \}'}{k_{nl} r} \right] \,,
\end{align}
with the prime denoting the derivative with respect to the dimensionless parameters $k_{nl} r$ or $q_{nl} r$, respectively, ${\cal N}_{nl}$ an overall normalization factor which is fixed by Eq.~\eqref{eq:norm}, and
\begin{align}
\beta_1(z) = \partial_z\left( \frac{j_l(z)}{z} \right)  \,, \quad \beta_3(z) = \frac{1}{2} \partial_z^2 j_l(z) + \left( \frac{1}{2} l (l + 1) - 1 \right)  \frac{j_l(z)}{z^2} \,.
\end{align}
The wave numbers $k_{nl}$ and $q_{nl}$ are also determined by the boundary condition. They do not depend on the angular mode number $m$, and for each $l$, the boundary conditions admit several solutions which are labeled by $n$ in increasing wave number. For CuAl6\% and the quadrupole mode $l=2$, the numerical values for the lowest modes are given in Tab.~\ref{tab:l2_eigenfrequencies}. For a sphere radius of $34$~cm as used in MiniGRAIL~\cite{Gottardi:2007zn}, this gives an eigenfrequency of 3.042~kHz for the lowest mode, which matches well with the measured value of 2.98~kHz.
It should be noted that, in practice, MiniGRAIL observed $\mathcal{O}(1\%)$ splittings that lifted the frequency degeneracy among different $m$ modes~\cite{Gottardi:2007zn}, demonstrating deviations from the idealized mode calculation. Therefore, even our more detailed calculation should be regarded as a guideline, although such small deviations can be calibrated and are not expected to affect our sensitivity estimates.

\renewcommand{\arraystretch}{1.5}
\setlength{\tabcolsep}{5.5pt}
\begin{table}[h]
 \begin{tabular}{l | c  c c c c c c c c c }
    $n$ & 1 & 2 & 3 & 4 & 5 & 6 & 7 & 8 & 9 & 10  \\ \hline
   $k_{n2} R$ & 2.65 &  5.11 &  8.63 &  11.06 &  12.31 &  15.35 &  17.99 &  18.7&  21.75 & 24.58 \\
    $q_{n2} R$ & 1.31 &  2.53 &  4.28 &  5.49 &  6.10 &  7.61 &  8.92 &  9.27 &  10.78 &  12.18\\
    $f_{n2}\times  (R/m)$~[kHz] & 1.03 &  1.99 & 3.37 &  4.32 &  4.8 &  5.99 &  7.02 &  7.3 &  8.49 &  9.59 \\
$ {\cal N}^{-1}_{n2} \times  10^3$ & 3.54 &  0.09 & 2.02 &  2.43 & 0.8 &  0.2 &  0.63 &  0.27 &  0.04 &  0.22
 \end{tabular}
 \caption{Lowest 10 eigenfrequencies for the quadrupolar mode of a solid CuAl6\% sphere. The linear eigenfrequency is obtained as $f = \omega/(2 \pi)$. The last row gives the normalization factors.}
  \label{tab:l2_eigenfrequencies}
\end{table}

\begin{figure}[b]
 \centering
 \includegraphics[width = 0.6 \columnwidth]{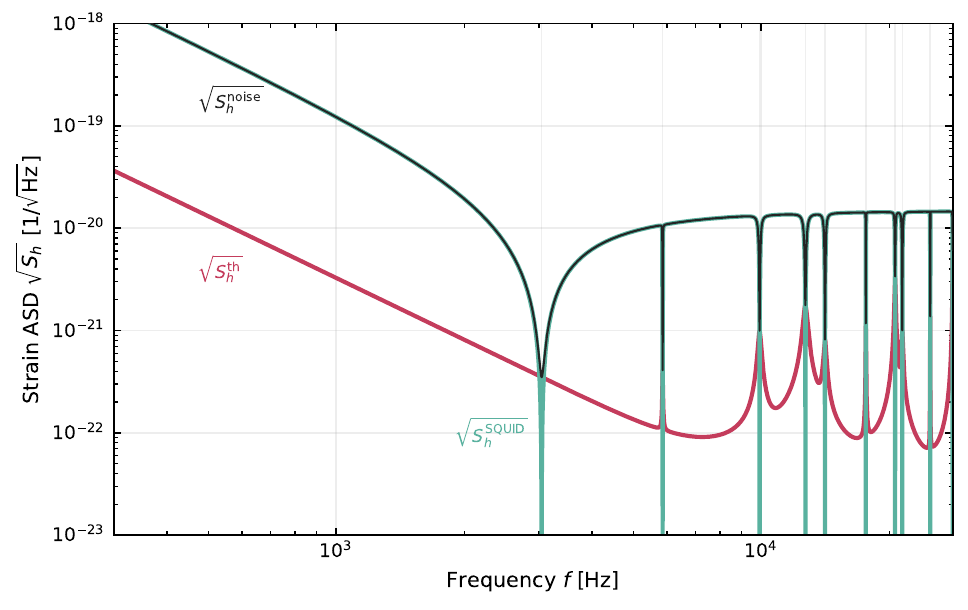}
 \caption{Strain-equivalent noise amplitude spectral density for the MiniGRAIL-inspired sphere with $R = 0.34$~m, $Q_n = Q = 10^7$ and a magnetic field gradient of $B'_r = 1~{\rm T/mm}$. The black curve is the total noise, while the solid red and green curves show the thermomechanical and SQUID contributions. The calculation includes the lowest ten $n22$ modes and uses their numerical eigenfrequencies, GW overlap coefficients, and signed area-integrated readout coefficients derived in this appendix, rather than the order-of-magnitude mode model used in Fig.~\ref{fig:sensitivity}. }
 \label{fig:sensitivity-minigrail}
\end{figure}

With these results at hand, one can immediately compute the overlap factors with the force induced by a passing GW. In particular, one can explicitly verify that the GW only couples to the $l=2$ mode and that for a fixed direction of the incoming GW, the overlap factor scales roughly as $1/n^2$. Concretely, the overlap factors for the first 10 modes with $l = 2, m = 2$ and a GW along the $z$ direction are given by
\begin{align}
 |\eta_{n22}/0.1|^2 = 9.0, \,0.93, \,0.032, \,0.0048, \,0.017, \,0.0041, \,0.00031, \,0.0031, \,0.0011, \,0.000037 \,.
\end{align}

We denote by $\alpha_{\bm n}(\theta,\phi)=\hat e_r\cdot\bm u_{\bm n}(R,\theta,\phi)$ the dimensionless radial surface profile of a mode. The measurable displacement-to-flux 
transfer function is obtained by integrating this profile over the instrumented area, weighted by the magnetic-field gradient and relative winding signs, giving the factor $\langle A\alpha_{\bm n}\partial_rB_r\rangle$. For the simple signed four-quadrant readout which sums over the sphere surface with a prefactor ${\rm sign}(\partial_rB_r \cos(2\phi))$, and approximating the radial magnetic gradient amplitude as fixed, the readout factor is obtained by
\begin{align}
\frac{\langle A\alpha_{n22}\partial_rB_r\rangle}{R^2|\partial_rB_r|} & =4\times \int_0^\pi \sin \theta d\theta \int_{-\pi/4}^{\pi/4}d\phi \,\hat e_r \cdot \bm u_{n22}(R) \\
& \simeq 4 \times\{-1.49,\, 0.02, \,-0.48, \,1.52, \,-0.53, \,0.11,\, -1.41, \,0.62, \,-0.04,\, 1.33\} \,.
\end{align}
The mean magnitude of the one-quadrant integrals over the first ten modes is approximately $0.34\pi/\sqrt{2}$. For $A=4\pi R^2/\sqrt{2}$, we therefore find $|\langle A\alpha_{n22}\partial_rB_r\rangle|/(A\partial_rB_r)\simeq0.34$, motivating the approximation $\langle A\alpha_{n22}\partial_rB_r\rangle=0.3\,A B'_r$ in the simplified model in the main text. The numerical calculation in this appendix instead uses the mode-dependent signed integrals.

For the quality factor, we will assume, as in the main text, $Q_{n22} \simeq Q = 10^7$. This is expected to be a good approximation for the first few modes considered here, though the quality factor is expected to drop for even higher modes, typically as $1/\omega$~\cite{Zener:1937zz,Gue:2026kga}.

We now have all the information needed to estimate the sensitivity of a spherical detector based on the MiniGRAIL parameters~\cite{Gottardi:2007zn}, using the numerical coefficients derived from the respective mode functions. The result is shown in Fig.~\ref{fig:sensitivity-minigrail}.
Notably, the sensitivity around the first resonance is $S_h^\text{noise} \simeq 4.1 \cdot 10^{-22}/\sqrt{\text{Hz}}$, limited by thermomechanical noise. This is similar to the sensitivity achieved by MiniGRAIL, $S_h^\text{noise} \simeq 1.5 \cdot 10^{-20}/\sqrt{\text{Hz}}$, which in this regime was limited by the thermomechanical noise of the transducers~\cite{Gottardi:2007zn}. While a dedicated comparison requires detailed knowledge about the experimental setup, it is reassuring to find that similar sensitivities have been reached in comparable setups.

\section{Halbach array and magnetic readout}
\label{app:halbach}

An ideal planar Halbach array is made of a large amount of long magnetized objects of finite width, whose magnetization direction rotates between adjacent magnets. This enhances the magnetic field on one side of the array while suppressing it on the other~\cite{Mallinson:1973,Halbach:1980,Blumler:2023qvb}. For a continuously rotating magnetization with period $\lambda=2\pi/k$ and magnet thickness $t$, the field above the enhanced side is~\cite{Blumler:2023qvb}
\begin{align}
 \bm B(x,z)=B_R\left(1-e^{-kt}\right)e^{-kz}
 \left[\sin(kx)\,\hat{\bm e}_x+\cos(kx)\,\hat{\bm e}_z\right]\,,
 \label{eq:planar-Halbach}
\end{align}
where $x$ and $z$ are respectively tangential and normal to the array. {The array is shift-symmetric in the $y$-direction.} While we work with a magnet which surrounds a sphere, the curvature radius is much larger than the period length, and so locally, we have  $B\sim B_0 (\cos(kR\phi )\hat{\phi}+\sin(kR\phi)\hat{r})e^{-k(R-r)}$. Note that this implies that to achieve a large field, the distance between the sphere and magnet must be extremely small, something that might lead to engineering challenges in terms of assembly. We further for simplicity take the gradient to be $\partial_r B_r\sim k B_0 {\rm sign}(\sin(kR\phi))\equiv B_0 w(\phi)$.

The orientation of neighboring pickup loops first follows the sign of $w(\phi)$, so their displacement-induced fluxes add despite the alternating gradient. On angular scales large compared with one array period, each circuit $a$ can additionally implement a piecewise-constant signed weighting $w_a(\theta,\phi)$. In the step-function approximation, its static and mode-induced fluxes scale as
\begin{align}
 \Phi_a^{(0)}&\propto B_0\int d\Omega\,w_a(\theta,\phi)\,,
 &
 \delta\Phi_{a,\bm n}&\propto B'_r c_{\bm n}
 \int d\Omega\,w_a(\theta,\phi)\alpha_{\bm n}(\theta,\phi)\,.
\end{align}
Since $\int d\Omega\,Y_{lm}=0$ for every $(l,m)\neq(0,0)$, a weighting matched to any quadrupolar mode can have zero mean, cancelling the leading static flux while retaining a nonzero mode response. The printed-loop segments with the signed four-quadrant readout provide a piecewise approximation to these weightings.

The pole-to-pole meridional loops shown in Fig.~\ref{fig:setup} are only a schematic $m=\pm2$ readout as their polar integral vanishes for the $l=2,|m|\neq 2$. Segmenting the traces in latitude gives additional independent weightings. With sufficiently many independently read segments, one can construct five zero-mean circuits whose response matrix is full rank over the five real $l=2$ modes. We leave the optimized layout, together with corrections from tangential motion, changes of loop orientation or area, and the finite-cell magnetic field, to a dedicated experimental design.

The inductance $L_p$ of an array of narrow loops with alternating orientation can be estimated as follows. Given a signal current $I_\text{sig}$, the associated energy is given by $U = \tfrac{1}{2} L_p I_\text{sig}^2$. With usual expression for a magnetic field generated by a single wire at a distance $r$, $B = I/(2 \pi r)$, we can equivalently estimate the the energy per unit length of wire as
\begin{align}
 \frac{dU}{dy} = \int_d^\lambda dr \, 2 \pi r \frac{B^2}{2} = \int_d^\lambda dr \frac{I_\text{sig}^2}{4 \pi} \ln(\lambda/d) \,,
\end{align}
where the radial integration runs from the wire diameter $d$ to the characteristic distance between two neighbouring wires $\lambda$. The total energy is obtained by multiplying with the typical wire length of a single lapse $2 R$ and the number of lapses needed, $2 \pi R/\lambda$, yielding $U = R I_\text{sig}^2/(4 \pi) \ln(\lambda/d)$. Equating this with the expression above gives
\begin{align}
 L_p = \frac{2 R^2}{\lambda} \ln(\lambda/d) \sim R^2/\lambda \,,
 \label{eq:inductance}
\end{align}
which roughly corresponds to the length of the total wiring used. From this we can estimate the SQUID coupling as
\begin{align}
 \kappa = \frac{\alpha}{2} \sqrt{\frac{L}{L_p}} \simeq 2 \cdot 10^{-3} \,,
\end{align}
where we have assumed $L \simeq 1$~nH for the intrinsic inductance of the SQUID, $\alpha \simeq 1/\sqrt{2}$~\cite{Kahn:2016aff} as well as benchmark values $R = 0.34$~m and $\lambda = 1$~mm. We note that by dividing the same wiring among several circuits, we likely increase $\kappa$ somewhat, but as each circuit would have its own SQUID noise, the effective SQUID noise is also increased to an extent. For simplicity, we simply take the above $\kappa^{-2}$ as an estimate to the effective enhancement of the readout noise in strain-equivalent units.
A reliable determination of both the electrical resonances and the coupling requires the full mutual-inductance and capacitance calculation of the system, including the possibility of coupling to the resonant sphere should it be conducting, as it was for MiniGRAIL. We leave such calculations to a future experimental design.

Finally, we note that Eq.~\eqref{eq:inductance} implies $S_\Phi^\text{SQUID} \propto \kappa^{-2} \propto R^{2}$, and hence an increase of the SQUID noise for larger sphere radii. However, the PSD signal transfer function scales as $A^2 \sim R^4$, so that the signal to squid noise ratio scales as $R^2$. Larger spheres thus allow to suppress the SQUID noise, even taking into account the penalty from the inductance.

\section{Including additional thermomechanical noise modes}
\label{app:noise}
\begin{figure}[t]
 \centering
 \includegraphics[width = 0.48 \columnwidth]{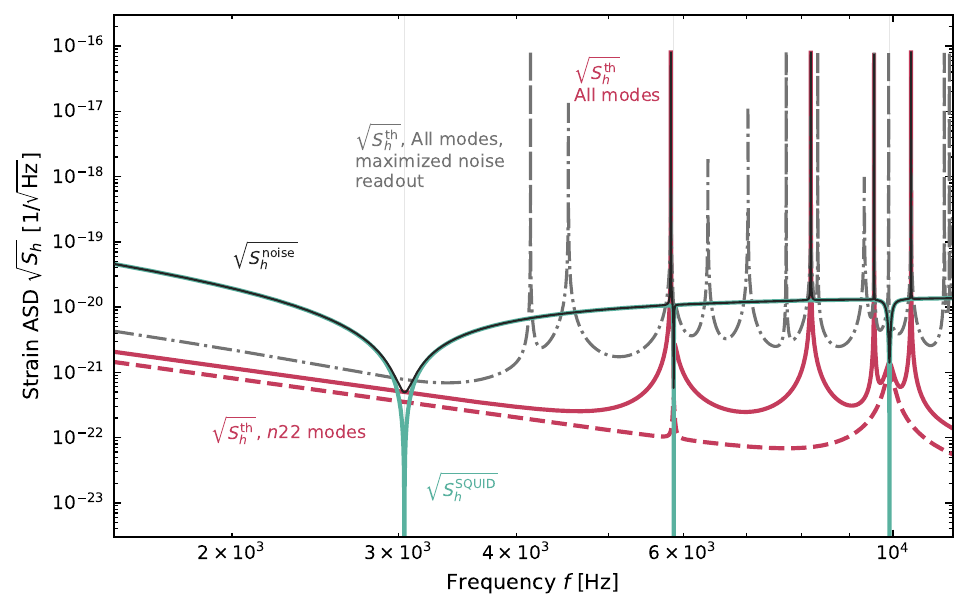} \hfill
 \includegraphics[width = 0.48 \columnwidth]{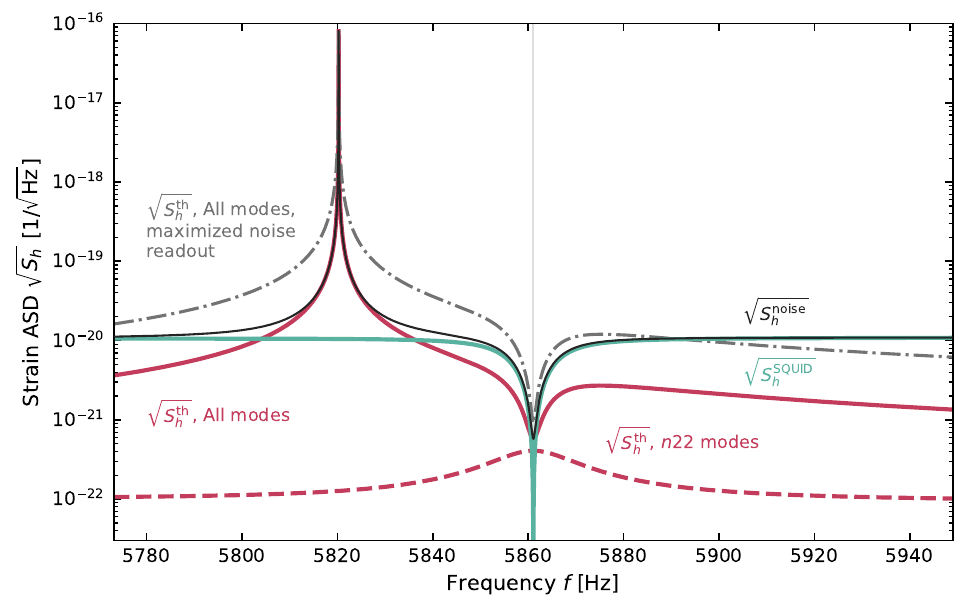}
 \caption{Thermomechanical-noise estimate including all spheroidal resonances below 12~kHz. The solid red curve uses the signed four-quadrant readout, while the dashed red curve retains only the $n22$ thermal modes. The gray dash-dotted curve is shown only for comparison and, for each noise mode separately, uses a spatial readout pattern matched to the sign of that mode's surface displacement, thereby artificially maximizing its contribution to the ASD. The total noise and SQUID contribution are shown in black and green. Vertical lines mark the first three $n22$ signal resonances. The left panel gives the broadband view, and the right panel zooms in on the $222$ signal resonance near 6~kHz and the nearby non-signal $l=4$ thermomechanical resonance.}
 \label{fig:sensitivity-noise-all-modes}
\end{figure}

The estimate above assumes that spectral and spatial information can be used to isolate the $l=2$, $m=2$ signal modes. Here we retain the lowest three $n22$ modes in the signal response but include all 150 (at 18 resonance frequencies) spheroidal modes whose resonances lie in the plotted range below 12~kHz when computing the thermomechanical noise:
\begin{align}
 S_\Phi^\text{sig} = \sum_{n=1}^3 (S_\Phi^\text{sig})_{n22} \,, \qquad S_\Phi^\text{th.mech} = \sum_{(n,l,m)\in {\cal M}_{<12\,{\rm kHz}}} (S_\Phi^\text{th.mech})_{nlm} \,.
\end{align}
We again use the MiniGRAIL-inspired parameters $R = 0.34$~m, $Q_{\bm n} = Q = 10^7$, as well as a magnetic field gradient $B'_r = 1~{\rm T/mm}$. For every mode, the radial readout coefficient is evaluated with the same fixed four-quadrant pattern, proportional to $\int d\Omega\,\operatorname{sign}[\cos(2\phi)]Y_{lm}(\theta,\phi)$, and independent members of each $m$ multiplet are summed in power. Figure~\ref{fig:sensitivity-noise-all-modes} compares this fixed-readout result with the $n22$-only estimate and with an artificially maximized, mode-dependent noise readout.

Since the $122$ mode is the lowest mode in the spectrum, with some significant separation in frequency from the next modes, the impact of the additional noise modes is negligible around this resonance.
At larger frequencies, we find as before that the sensitivity is mostly limited by SQUID noise and not thermomechanical noise, and adding the full mode spectrum does not change this conclusion outside of the new resonant peaks. 
Targeting the rather broadband signals expected for most high frequency gravitational wave signals, one could map out and mask the narrow frequency windows corresponding to these peaks, without significantly impacting the sensitivity. In this sense it is a good approximation to take SQUID noise to be the dominant noise source for these parameters outside of the $n22$ resonance windows.

If one of these new peaks of the thermomechanical noise spectrum occurs very close to a (higher) signal resonance one might worry about a significantly enhanced thermomechanical noise contribution reducing the on-resonance sensitivity. This is illustrated in the right panel of Fig.~\ref{fig:sensitivity-noise-all-modes} by the $l=4$ thermomechanical resonance at $5.82$~kHz near the $222$ signal resonance at $5.86$~kHz. However, since the resonances are all very narrow, enhancement turns out to be very moderate, as can be seen by comparing the dashed and solid red lines around the signal resonance at $5.86$~kHz. This should be seen as a further proof of principle that the filtering of modes assumed in the main text can be plausibly achieved. In practice, this challenge becomes more difficult at larger frequencies. At this point, the pickup loop wiring must compromise between optimal mode separation and a minimal number of circuits.

\section{An illustrative toy-example of a GW burst in the time domain}
\label{app:time-domain}

In the frequency regime of interest, a common target are are short duration GW bursts, arising e.g.\ by mergers of light primordial black holes. This poses a challenge for GW detectors, as the resonator cannot coherently accumulate energy over many periods. On the other hand, the ring-down signal of the excellent mechanical resonator can still be used to detect these types of signals.

As a very simple toy-model, we will model the GW signal as a monochromatic signal during a finite duration $T \sim 1/\omega_g$,
\begin{align}
h^\Theta_A(t)=h_0^A e^{i\omega_g t}\Theta(t)\Theta(T-t)+h.c.\,.
 \label{eq:h}
\end{align}
{where for simplicity we take the amplitude $h_0^A$ to be real.}
We stress however that the sensitivity analysis in the main text does not depend on the choice of GW waveform.

\begin{figure}
 \centering
 \includegraphics[width =0.92\textwidth]{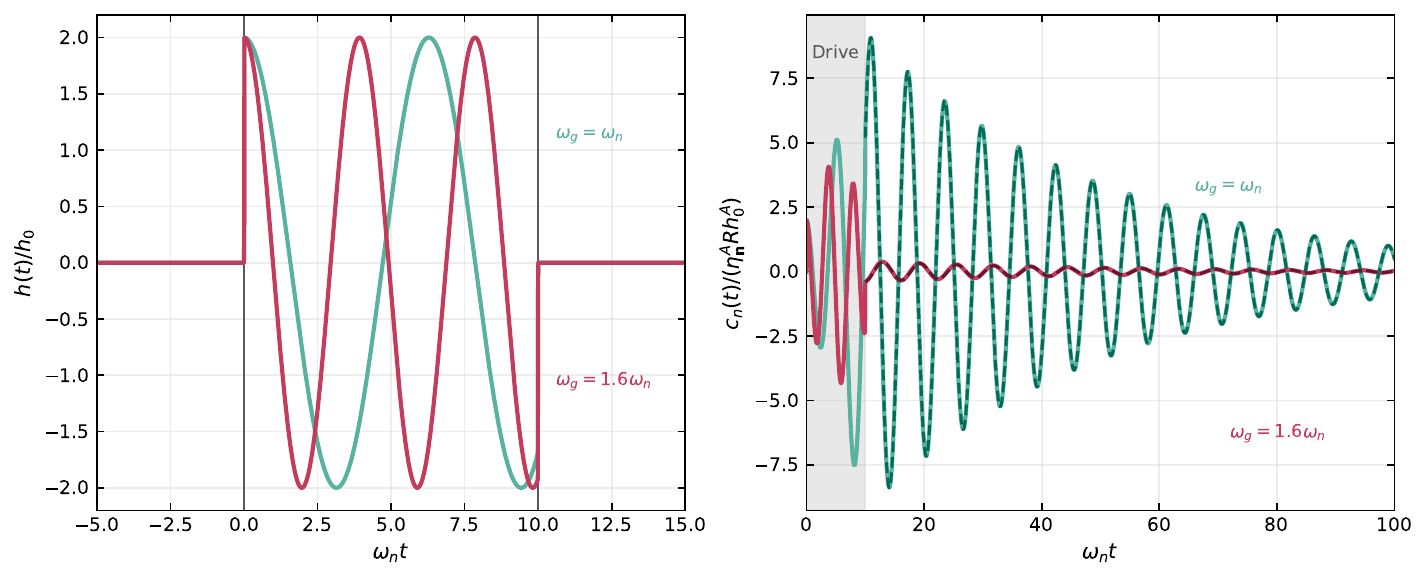}
 \caption{Finite-duration toy GW signals (left) and the resulting sphere response (right) for $Q_{\bm n} = 20$ (chosen so that the ring-down decay is visible on the plotted time scale). Green denotes the resonant case, $\omega_g=\omega_{\bm n}$, and red denotes the off-resonant case, $\omega_g=1.6\omega_{\bm n}$. The shaded region in the right panel marks the drive interval. During the subsequent ring down, the exact responses are closely approximated by the dashed curves from Eqs.~\eqref{eq:cnres} and \eqref{eq:cnoffres}. }
 \label{fig:cnt}
\end{figure}

To aid our intuition of the type of signals to search for in the data stream, it is useful to consider also the response of our setup in the time domain. Starting from Eq.~\eqref{eq:cn-general}, we can express the coefficients $c_{\bm n}(t)$ appearing in Eq.~\eqref{eq:u_ansatz} as
\begin{align}
c_{\bm n}(t)=\eta_{\bm n}^A R\int\frac{d\omega}{\sqrt{2\pi}}\widetilde{\ddot h}^{\Theta}_A(\omega)\tilde F(\omega)e^{-i\omega t} \,,
\end{align}
 with $\tilde F(\omega) = - (\omega^2 - \omega_{\bm n}^2 + i \omega \omega_{\bm n}/Q_{\bm n})^{-1}$. Using
\begin{align}
 \int \frac{d \omega}{\sqrt{2 \pi}} \tilde F(\omega) \tilde f(\omega) e^{- i \omega t} = \int \frac{dt'}{\sqrt{2 \pi}} f(t') F(t - t') \,,
 \label{eq:F-trick}
\end{align}
for any function $f(t)$ and
\begin{align}
 F(t) = \sqrt{2 \pi} \Theta(t) \frac{1}{\omega_d} e^{- \Gamma t/2} \sin (\omega_d t) \,,
\end{align}
with $\Gamma = \omega_{\bm n}/Q_{\bm n}$, $\omega_d^2 = \omega_{\bm n}^2 - (\Gamma/2)^2$ in the underdamped regime $Q_{\bm n} > 1/2$, we obtain
\begin{align}
c_{\bm n}(t)=\eta_{\bm n}^A R\int_{-\infty}^\infty\frac{dt'}{\sqrt{2\pi}}\ddot h_A^{\Theta}(t')F(t-t')\,.
\end{align}
The resulting time-domain response is shown in Fig.~\ref{fig:cnt}. The left panel shows two finite-duration toy GW signals: a resonant drive with $\omega_g=\omega_{\bm n}$ (green) and an off-resonant drive with $\omega_g=1.6\omega_{\bm n}$ (red). The right panel shows the corresponding sphere response, with the shaded interval indicating the duration of the drive. The slightly lighter dashed curves give the large-$Q$ ring-down approximations for $t>T$, and nearly perfectly match the exact calculation,
\begin{align}
 \frac{c_{\bm n}^\text{res}(t)}{\eta_{\bm n}^A R h_0^A} & \simeq \frac{1 }{2} e^{-\frac{t \omega_{\bm n}}{2 Q_{\bm n}}} \left(4 (Q_{\bm n} \sin (t \omega_{\bm n})+\cos (t \omega_{\bm n}))-e^{\frac{T \omega_{\bm n}}{2 Q_{\bm n}}} (4 Q_{\bm n} \sin (t \omega_{\bm n})+\cos (\omega_{\bm n} (t-2 T))+3 \cos (t \omega_{\bm n}))\right) \,,
 \label{eq:cnres} \\
 \frac{c_{\bm n}^\text{off}(t)}{\eta_{\bm n}^A R h_0^A} & \simeq \frac{2\, \omega_{\bm n}}{\omega_{\bm n}^2-\omega_g^2}  e^{-\frac{t \omega_{\bm n}}{2 Q_{\bm n}}} \left(e^{\frac{T \omega_{\bm n}}{2 Q_{\bm n}}} (\omega_g \sin (T \omega_g) \sin (\omega_{\bm n} (t-T))-\omega_{\bm n} \cos (T \omega_g) \cos (\omega_{\bm n} (t-T)))+\omega_{\bm n} \cos (t \omega_{\bm n})\right)\,.
 \label{eq:cnoffres}
\end{align}

From the above behavior, we can justify our focus on $t>T$. As mentioned in the main text, a transient signal, lasting $T\sim 1/\omega_g$ would lead to a response lasting $Q_n/\omega_n$ (with $\omega_n=\omega_1$ or $\omega_n\sim \omega_g$ depending on the regime, see App.~\ref{app:hf}). Hence focusing only on $t>T$, during which it is assumed the magnet motion is quickly damped ($Q_\text{magnet} \sim 1$), would lose at most $1/Q$ of the signal, and allows us to simplify our treatment by ignoring the GW effect on the magnet.

\section{High-frequency regime}
\label{app:hf}

It is instructive to consider the response function and noise contributions of our setup in the high-frequency limit. To obtain analytical expressions, we will use the simplified model for the modes of the resonant sphere employed in the main text, i.e.\ $\omega_{n22} = n \omega_1$, ${\eta_{n22}^\times=\eta/n^2}$, ${\langle A\alpha_{n22}\partial_rB_r\rangle= A\alpha B'_r}$, $Q_{n22} = Q$.

We start with the signal contribution
\begin{align}
S^\text{sig}_{x}(\omega)=\sum_n\frac{(\eta R/n^2)^2\omega^4}{(\omega^2-n^2\omega_1^2)^2+(n\omega_1\omega/Q)^2}S_h(\omega)\,.
\end{align}
For any given $\omega$, split the modes into ${n\omega_1<\omega}$ and ${n\omega_1>\omega}$, for which we can approximate
\begin{align}
\frac{\omega^4}{(\omega^2-n^2\omega_1^2)^2+(n\omega_1\omega/Q)^2}\simeq
\begin{cases}
1 & n\omega_1<\omega\\ {}
[\omega/(n\omega_1)]^4 & n\omega_1>\omega
\end{cases}\,.
\label{eq:scalings}
\end{align}
We see that the dominant contributions come from the first case, ${n\omega_1<\omega}$,
\begin{align}
S^\text{sig}_{x}(\omega)\big|_\text{HF}\simeq\sum_{n=1}^{\omega/\omega_1}(\eta R)^2n^{-4}S_h(\omega)\simeq(\eta R)^2S_h(\omega)\,.
 \label{eq:Ssig-HF}
\end{align}
To obtain the signal contribution to the flux PSD we need to multiply with the transfer function $\langle A\alpha B'_r \rangle^2$, which adds no new frequency dependence.

For comparison, also the free-falling limit commonly employed in the high-frequency regime, $\delta  x_i^\text{FF} = \frac{1}{2} h_{ij} x^j$, yields a scale invariant displacement PSD. We recover this from Eq.~\eqref{eq:Ssig-HF}.
The scaling ${\eta_{n22}\sim\eta/n^2}$ ensures that the lowest mode $(n=1)$ dominates. At $\omega \gg \omega_1$, the $\omega^2$ scaling of the GW force is compensated by the $1/\omega^2$ in the response of an oscillator driven with a frequency far above its eigenfrequency, resulting in a flat signal PSD at high frequencies, corresponding to the elastic free falling regime discussed in Ref.~\cite{Gue:2026kga}.

In Eq.~\eqref{eq:Ssig-HF} we have dropped the on-resonant contribution ${\omega\simeq n\omega_1}$, for which Eq.~\eqref{eq:scalings} evaluates to $Q^2$, and contributes as
\begin{align}
S^\text{sig}_{x}(\omega)\big|_\text{res}\simeq(\eta R)^2(Q^2/n^4)S_h(\omega)
 \label{eq:Ssig-res}
\end{align}
to the signal PSD. Comparing to Eq.~\eqref{eq:Ssig-HF}, we see that this contribution is subdominant for $\omega > \sqrt{Q} \omega_1$.

Next, let us turn to the thermomechanical noise, considering only the modes also populated by the signal (fixed $lm = 22$),
\begin{align}
S_{\Phi,n}^\text{th.mech.}(\omega)=(A\alpha B'_r)^2
\frac{2Tn\omega_1}{QM[(\omega^2-n^2\omega_1^2)^2+(n\omega_1\omega/Q)^2]}\,.
\label{eq:Sthmech-app}
\end{align}
In this case, the higher modes are no longer suppressed by the overlap $\eta_{n22}\sim {\eta/n^2}$, so at fixed $\omega$ the PSD is dominated by modes with ${n\omega_1\sim\omega}$. Evaluating Eq.~\eqref{eq:Sthmech-app} on resonance gives
\begin{align}
S_{\Phi}^\text{th.mech.}(\omega)\big|_\text{res}\simeq
(A\alpha B'_r)^2\frac{2TQ}{M\omega^3}\,.
 \label{eq:Sthmech-res}
\end{align}
At very high frequencies ${\omega>Q\omega_1}$ the modes are strongly overlapping, so that we can express the sum over modes as an integral,
\begin{align}
\int_0^\infty dn\,S_{\Phi,n}^{\rm th.mech.}(\omega)\big|_\text{HF}
&=(A\alpha B'_r)^2\frac{2T}{MQ}
\int_0^\infty dn\,\frac{n\omega_1}{(\omega^2-n^2\omega_1^2)^2+(n\omega_1\omega/Q)^2}\\
&\simeq(A\alpha B'_r)^2\frac{\pi T}{M\omega_1\omega^2}\,,
\end{align}
for $Q \gg 1$. This is however beyond our frequency range of interest. We thus use Eq.~\eqref{eq:Sthmech-res}, which drops as $\omega^{-3}$ at high frequencies, indicating that SQUID noise will eventually take over. Concretely, recalling  $S_\Phi^\text{SQUID}(\omega)=\kappa^{-2}10^{-12}\Phi_0^2/\text{Hz}$,
and using the typical numbers of the main text,
$R=0.34~{\rm m}$, $A=4\pi R^2/\sqrt{2}$, ${\alpha=0.3}$,
$B'_r=1~{\rm T/mm}$, $M=1.3\times10^3~{\rm kg}$,
and $\kappa=0.002$, we find 
\begin{align}
\frac{S_{\Phi,\mathrm{res}}^\text{th.mech.}}{S_\Phi^\text{SQUID}}
&\simeq
\left(\frac{T}{4.2~{\rm K}}\right)
\left(\frac{Q}{10^7}\right)
\left(\frac{\omega}{2\pi\times7~{\rm MHz}}\right)^{-3}
\left(\frac{M}{1.3\times10^3~{\rm kg}}\right)^{-1}
\left(\frac{\kappa}{0.002}\right)^2\nonumber\\
&\quad\times
\left(\frac{\alpha}{0.3}\right)^2
\left(\frac{A}{4\pi(0.34~{\rm m})^2/\sqrt{2}}\right)^2
\left(\frac{B'_r}{{\rm T/mm}}\right)^2.
\label{eq:th-squid-ratio-hf}
\end{align}
If $Q=10^7$ remained constant, Eq.~\eqref{eq:th-squid-ratio-hf} would give a crossover near $7~{\rm MHz}$. If instead $Q$ drops by $\sim2$ orders of magnitude for the high-frequency modes, the crossover occurs earlier, at a frequency of order $\sim {\rm MHz}$. {In this high-frequency regime (but still assuming $\omega_g \gg R$), the strain equivalent noise then reads
\begin{align}
S_h^\text{noise}(f>\text{MHz})=\frac{S_\Phi^\text{SQUID}S_h}{(A\alpha B'_r)^2S_x^\text{sig}|_\text{HF}}
\simeq\frac{\kappa^{-2}10^{-12}\Phi_0^2/\text{Hz}}{(A\alpha B'_r\eta R)^2}\,.
\end{align}
We note that once SQUID noise dominated, the strain-equivalent noise curve becomes $\omega$ independent, and we recover the high-frequency broadband regime claimed in Ref.~\cite{Domcke:2024mfu}. This is also the relevant expression to estimate the off-resonance sensitivity at lower frequencies, as indicated by the uppermost dashed gray line in Fig.~\ref{fig:sensitivity}.}

To estimate the strain-equivalent noise in the regime up to a MHz (for which $\omega < \sqrt{Q} \omega_1$), we instead use Eqs~\eqref{eq:Ssig-res} and \eqref{eq:Sthmech-res},
 \begin{align}
S_h^\text{noise}(100~\text{kHz}<f<\text{MHz})=\frac{S_\Phi^\text{th.mech.}}{(A\alpha B'_r)^2S_x^\text{sig}}\bigg|_\text{res}S_h
\simeq\frac{2T\omega}{(\eta R\omega_1^2)^2QM}\,.
 \end{align}
This is the estimate we show in dashed gray for the thermomechanical noise contribution in Fig.~\ref{fig:sensitivity} in the main text at intermediate frequencies.

\end{document}